\documentclass[aps,prb,superscriptaddress,longbibliography,twocolumn]{revtex4-2}

\usepackage{graphicx}
\usepackage{bm}
\usepackage{hyperref}
\usepackage{todonotes}
\usepackage{color}
\usepackage{comment}
\usepackage{verbatim}
\usepackage{soul}
\usepackage{placeins}

\begin{document}

\def\cuiro{$\mathrm{Cu_2IrO_3}$}
\def\aliiro{$\mathrm{\alpha{\text -}Li_2IrO_3}$}

\title{Complex pressure-temperature structural phase diagram of honeycomb iridate Cu$_2$IrO$_3$}

\author{G. Fabbris}
\email{gfabbris@anl.gov}
\affiliation{Advanced Photon Source, Argonne National Laboratory, Argonne, Illinois 60439, USA}

\author{A. Thorn}
\affiliation{Department of Physics, Applied Physics and Astronomy, Binghamton University, State University of New York, PO Box 6000, Binghamton, New York 13902-6000, USA}

\author{W. Bi}
\affiliation{Advanced Photon Source, Argonne National Laboratory, Argonne, Illinois 60439, USA}
\affiliation{Department of Geology, University of Illinois at Urbana-Champaign, Urbana, Illinois 61801, USA}
\affiliation{Department of Physics, University of Alabama at Birmingham, Birmingham, AL 35294, USA}

\author{M. Abramchuk}
\affiliation{Physics Department, Boston College, Chestnut Hill, Massachusetts 02467, USA}

\author{F. Bahrami}
\affiliation{Physics Department, Boston College, Chestnut Hill, Massachusetts 02467, USA}

\author{J. H. Kim}
\affiliation{Advanced Photon Source, Argonne National Laboratory, Argonne, Illinois 60439, USA}

\author{T. Shinmei}
\affiliation{Geodynamics Research Center, Ehime University, Matsuyama, 790-8577, Japan}

\author{T. Irifune}
\affiliation{Geodynamics Research Center, Ehime University, Matsuyama, 790-8577, Japan}
\affiliation{Earth-Life Science Institute, Tokyo Institute of Technology, Tokyo, Japan}

\author{F. Tafti}
\affiliation{Physics Department, Boston College, Chestnut Hill, Massachusetts 02467, USA}

\author{A. N. Kolmogorov}
\affiliation{Department of Physics, Applied Physics and Astronomy, Binghamton University, State University of New York, PO Box 6000, Binghamton, New York 13902-6000, USA}

\author{D. Haskel}
\affiliation{Advanced Photon Source, Argonne National Laboratory, Argonne, Illinois 60439, USA}

\date{\today}

\begin{abstract}

\ \ \cuiro\ is among the newest layered honeycomb iridates and a promising candidate to harbor a Kitaev quantum spin liquid state. Here, we investigate the pressure and temperature dependence of its structure through a combination of powder x-ray diffraction and x-ray absorption fine structure measurements, as well as \emph{ab-initio} evolutionary structure search. At ambient pressure, we revise the previously proposed $C2/c$ solution with a related but notably more stable $P2_1/c$ structure. Pressures below 8~GPa drive the formation of Ir-Ir dimers at both ambient and low temperatures, similar to the case of $\mathrm{Li_2IrO_3}$. At higher pressures, the structural evolution dramatically depends on temperature. A large discontinuous reduction of the Ir honeycomb interplanar distance is observed around 15~GPa at room temperature, likely driven by a collapse of the O-Cu-O dumbbells. At 15~K, pressures beyond 20~GPa first lead to an intermediate phase featuring a continuous reduction of the interplanar distance, which then collapses at 30~GPa across yet another phase transition. However, the resulting structure around 40~GPa is not the same at room and low temperatures. Remarkably, the reduction in interplanar distance leads to an apparent healing of the stacking faults at room temperature, but not at 15~K. Possible implications on the evolution of electronic structure of \cuiro\ with pressure are discussed.

\end{abstract}

\maketitle

\section{Introduction}

The prediction that a Kitaev quantum spin liquid (QSL) state may emerge in honeycomb systems with strong spin orbit coupling (SOC) has triggered an intense activity around $4d$ and $5d$ materials due to exciting prospects for topological quantum computing \cite{Kitaev2006, Jackeli2009}. However, realizing a Kitaev QSL in these systems has proven to be challenging, because even small structural distortions enhance Heisenberg interactions that compete with the Kitaev exchange, and ultimately lead to undesired long-range magnetic order. Applying high pressure is a natural choice in an attempt to tune the structure and hence magnetic interactions. However, the presence of edge shared IrO$_6$ octahedra result in relatively short Ir-Ir bonds that are prone to dimerization, as observed in $\mathrm{Li_2IrO_3}$ polymorphs \cite{Veiga2017, Veiga2019, Hermann2018, Clancy2018}, which destroys the $J_{\mathrm{eff}} = 1/2$ state and related bond directional exchange anisotropy.

\cuiro\ stands out as one of the recently discovered honeycomb iridates for which there is evidence for a Kitaev QSL state \cite{Abramchuk2017, Choi2019, Takahashi2019, Kenney2019, Pal2020}, that also include $\mathrm{H_2LiIr_2O_6}$ and $\mathrm{Ag_2LiIr_2O_6}$ \cite{Kitagawa2018, Bahrami2019}. Interestingly, the key feature uniting these compounds is a larger separation between the honeycomb layers, realized in \cuiro\ by the presence of O-Cu-O dumbbell bonds, which reduces dimensionality and suppresses magnetic order. However, separating the layers also leads to stacking faults \cite{Abramchuk2018}. Such disorder affects the interpretation of thermodynamic measurements, posing a challenge in determining the nature of their ground state \cite{Knolle2019, Kao2021, Bahrami2021}. \cuiro\ has the additional feature that the (nominal) Cu$^{1+}$ ions within the honeycomb layer are in an octahedral environment, which is expected to favor the 2+ valence state. Indeed, spectroscopic data suggests that about 1/2 to 1/3 of the in-plane Cu are 2+ \cite{Kenney2019}. The 3:1 occupancy ratio of Cu ions in dumbbells versus honeycomb layers results in an average Cu oxidation state of $\sim 1.1+$. Interestingly, while Ir is expected to receive this extra charge, ambient pressure x-ray absorption at the Ir $L_{3,2}$ edges yields $\langle L.S \rangle = 2.85(6)$ (in units of $\hbar^2$), which indicates that the $J_{\mathrm{eff}} = 1/2$ picture is still valid \cite{Laguna-Marco2010, Haskel2012}. However, the relatively large fraction of magnetic Cu ions in the honeycomb plane likely disrupts the Ir-Ir magnetic exchange interactions, potentially competing with the Kitaev exchange.

Here we explore the structural phase diagram of \cuiro\ at high pressure. A series of pressure-induced phase transitions are observed in powder x-ray diffraction (PXRD) measurements at room temperature (RT) and 15 K (LT), as well as predicted with density functional theory (DFT)-based evolutionary structure searches. According to the global structure optimization at ambient pressure, the \cuiro\ ground state has the $P2_1/c$ symmetry rather than the previously proposed $C2/c$ \cite{Abramchuk2017} or its refined $C2/m$ derivative. The three monoclinic phases have similar morphologies and cannot be uniquely distinguished using PXRD, but the proposed $P2_1/c$ solution is significantly more stable in DFT. Upon pressurization, at both ambient and low temperature the first transition occurs at about 7~GPa into a structure with $P\overline{1}$ symmetry. DFT and x-ray absorption fine structure (XAFS) measurements at the Ir $L_3$ edge reveal that such $P\overline{1}$ phase is composed of Ir-Ir dimers, similar to those observed in $\mathrm{Li_2IrO_3}$ \cite{Hermann2018, Veiga2019}. Higher pressures up to 40 GPa lead to a single phase transition at RT around 15 GPa, but two transitions occur at LT around 20 and 30 GPa. While both RT and LT structures around 40 GPa are ultimately characterized by a collapsed interplanar distance, significantly higher pressure is needed at LT to reach the collapsed phase ($\sim 30$~GPa, compared to $\sim 15$~GPa at RT), the resulting distances are quite different [$\mathrm{(d^{RT}-d^{LT})/d^{LT} \sim 5.5\%}$], and the out-of-plane compressibility is significantly \emph{larger} at LT. Surprisingly, the stacking faults appear to be mostly healed at room temperature, while still present at 15~K. This implies that the combination of high pressure and elevated temperature leads to a larger structural correlation between honeycomb planes, suggesting that high-pressure annealing might lead to a well ordered structure. We use the DFT results to discuss the consequences of the observed transitions to the \cuiro\ electronic structure. Finally, we expect that this work will not only stimulate further investigations on the electronic properties of \cuiro\ at high pressures, but also motivate similar studies on the other honeycomb iridates with intercalated layers to verify the generality of the structural phase transitions observed here.

\section{Methods}

\subsection{Powder x-ray diffraction}
Powder x-ray diffraction was collected in angle dispersive mode at 11-BM (ambient pressure) and HPCAT 16-BM-D (high pressure) beamlines of the Advanced Photon Source, Argonne National Laboratory. Ambient pressure data were measured using $\lambda = 0.4581 ~ \mathrm{\AA}$ ($E \approx 27 ~ \mathrm{keV}$). A kapton capillary of 0.5 mm diameter was coated with the sample using vacuum grease, and placed inside another kapton capilary of 0.8 mm inner diameter. The sample was spun at high frequency ($> 60$ Hz) to improve powder averaging, and data collected using a set of 12 independent Si(111) analyzers \cite{Lee2008}. High pressure was generated using Princeton-type symmetric cells fitted with diamond anvils of 300 $\mu$m culet diameter and Re gaskets. The x-ray wavelength was set to $\lambda = 0.4134 ~ \mathrm{\AA}$ ($E \approx 30 ~ \mathrm{keV}$), which increases the accessible reciprocal space (pressure cell aperture $\approx 20^{\circ}$ in $2\theta$), and minimizes the absorption from the diamond anvils and boron carbide seat. Both ruby fluorescence and gold lattice constant served as manometers \cite{Holzapfel2001, Ragan1992, Dewaele2008}. Low temperature (15~K) measurements were performed using a He flow cryostat, with pressure applied in-situ. Data were also collected in a separate measurement at room temperature to avoid the background generated by the cryostat windows. Helium gas was used as pressure medium, and was loaded using the GSECARS gas loading facility \cite{Rivers2008}. The 2D images from the MAR3450 detector were converted to 1D diffractograms using the Dioptas software \cite{Supplemental,Prescher2015}, which was also used to mask diamond Bragg peaks, as well as correct for the diamond and seat absorption. Rietveld refinements were done using GSASII \cite{Toby2013}.

\subsection{X-ray absorption fine structure}
High pressure isothermal x-ray absorption fine structure (XAFS) measurements were performed at 15 and 300 K at the 4-ID-D beamline of the Advanced Photon Source, Argonne National Laboratory. Temperature was controlled using a He flow cryostat. Data were collected using a diamond anvil cell fitted with nanopolycrystalline (NPD) diamond anvils of 400 $\mu$m culet diameter and stainless steel gasket \cite{Irifune2003, Ishimatsu2012}. Neon was used as pressure medium \cite{Rivers2008}, and ruby fluorescence as manometer \cite{Ragan1992, Dewaele2008}. Incident and transmitted x-rays were detected using N$_2$ and Ar filled ion chambers, respectively. The data were normalized and modelled using the Larch package and FEFF8 \cite{Newville2013, Ankudinov1998}. XAFS temperature dependence was collected at ambient pressure, from which the E$_0$ (4.64 eV) and S$_0^2$ (0.98) parameters of the XAFS equation were extracted, and used to reduce the number of variables in the modelling of the high pressure data (see Supplementary Material for further details \cite{Supplemental}).

\subsection{\emph{Ab-initio} structure search}

All DFT calculations were performed with {\small VASP} using the projector-augmented wave (PAW) potentials and the Perdew-Burke-Ernzerhof (PBE) exchange correlation functional in the generalized gradient approximation (GGA) \cite{Kresse1993, Kresse1996, Blochl1994, Perdew1996, Perdew1997}. A 500-eV plane-wave energy cutoff and $6\times 6\times 4$ or finer Monkhorst-Pack k-meshes ensured good numerical convergence. Final stability and band structure results were obtained in the DFT+U approximation with the SOC included \cite{Monkhorst1976, Pack1977, Anisimov1997,Dudarev1998}. We chose typical $U_{eff}$ values of 2 eV and 6.5 eV for Ir and Cu, respectively \cite{Sun2018, Anisimov1997}.

Unconstrained searches for lowest-enthalpy phases at the DFT+U level were driven by an evolutionary algorithm implemented in {\small MAISE} \cite{maise}. As in our previous joint studies \cite{AK23, Sun2018}, we chose not rely on any crystallographic input from high-pressure experiments for an independent identification of the ground states. Global optimization runs at 0, 8, 16, and 40 GPa with 12-16 members in the population were initialized with random structures for 12-atom unit cells or randomized $C2/c$ structure for 24-atom unit cells and evolved for up to 15 generations. The $C2/c$ seeds allowed us to accelerate the searches without constraining candidate structures to a particular symmetry or morphology. Due to the considerable computational cost of DFT calculations, the bulk of this investigation was conducted for pressures below 20 GPa.

\begin{figure}
\includegraphics[width = 8.3 cm]{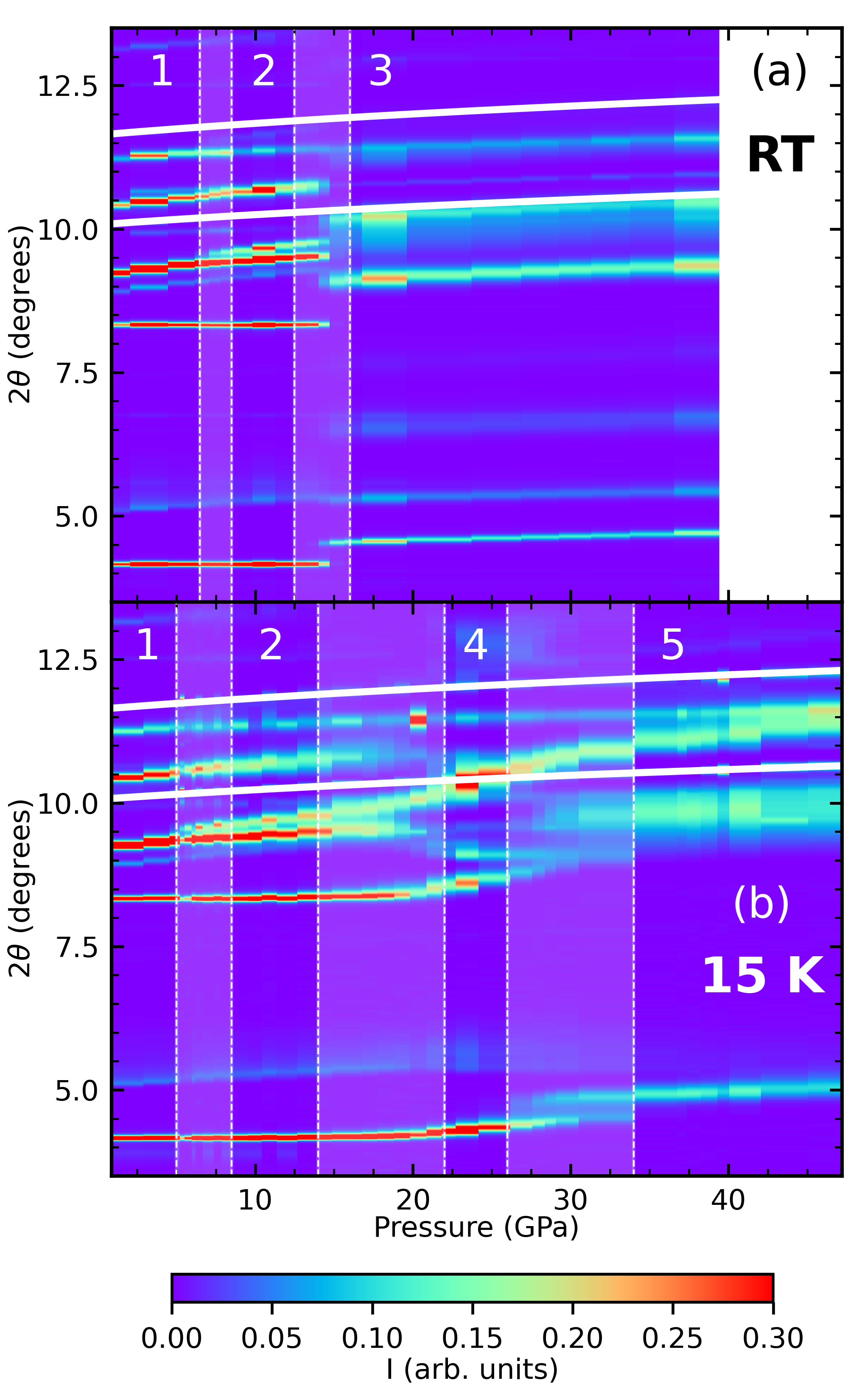}
\caption{Pressure dependence of the powder x-ray diffraction of \cuiro\ taken at (a) RT and (b) 15 K. Each temperature map was normalized to its maximum intensity, and peaks from Au manometer were masked. White dashed vertical lines and shaded area mark the coexistence regions that separate three distinct phases at room temperature (labeled here as 1, 2 and 3), and four at low temperature (1, 2, 4, and 5).}
\label{xrdexp}
\end{figure}

\section{Results}

Figure \ref{xrdexp} summarizes the pressure dependence of PXRD in \cuiro. Three phases are observed at room temperature below 40 GPa, which here are labeled 1, 2, and 3. The first phase transition happens at $7.5 \pm 1$ GPa, being marked by an apparent splitting of a few Bragg peaks. The onset of phase 3 occurs at $14.5 \pm 2$ GPa, and leads to a dramatic change in the diffraction pattern. The behavior of the Bragg peak at $2\theta \sim 4^{\circ}$ is particularly noteworthy as it is a measure of the interplanar distance. Its discontinuous increase in angular position is a clear signature that phase 3 features a collapsed interplanar distance. Phases 1 and 2 also occur at low temperature within similar pressure ranges, but pressures beyond 15 GPa drive \cuiro\ into different structures than those seen at RT. Interestingly, the interplanar distance is continuously reduced in phase 4, collapsing only at the onset of phase 5 around 30 GPa, about twice the pressure (15 GPa higher) than needed at RT. All observed phase transitions are reversible upon pressure release at RT \cite{Supplemental}. These features highlight a rich phase diagram in \cuiro, which we address in more detail in the following sections.

\begin{figure}
\includegraphics[width = 8.3 cm]{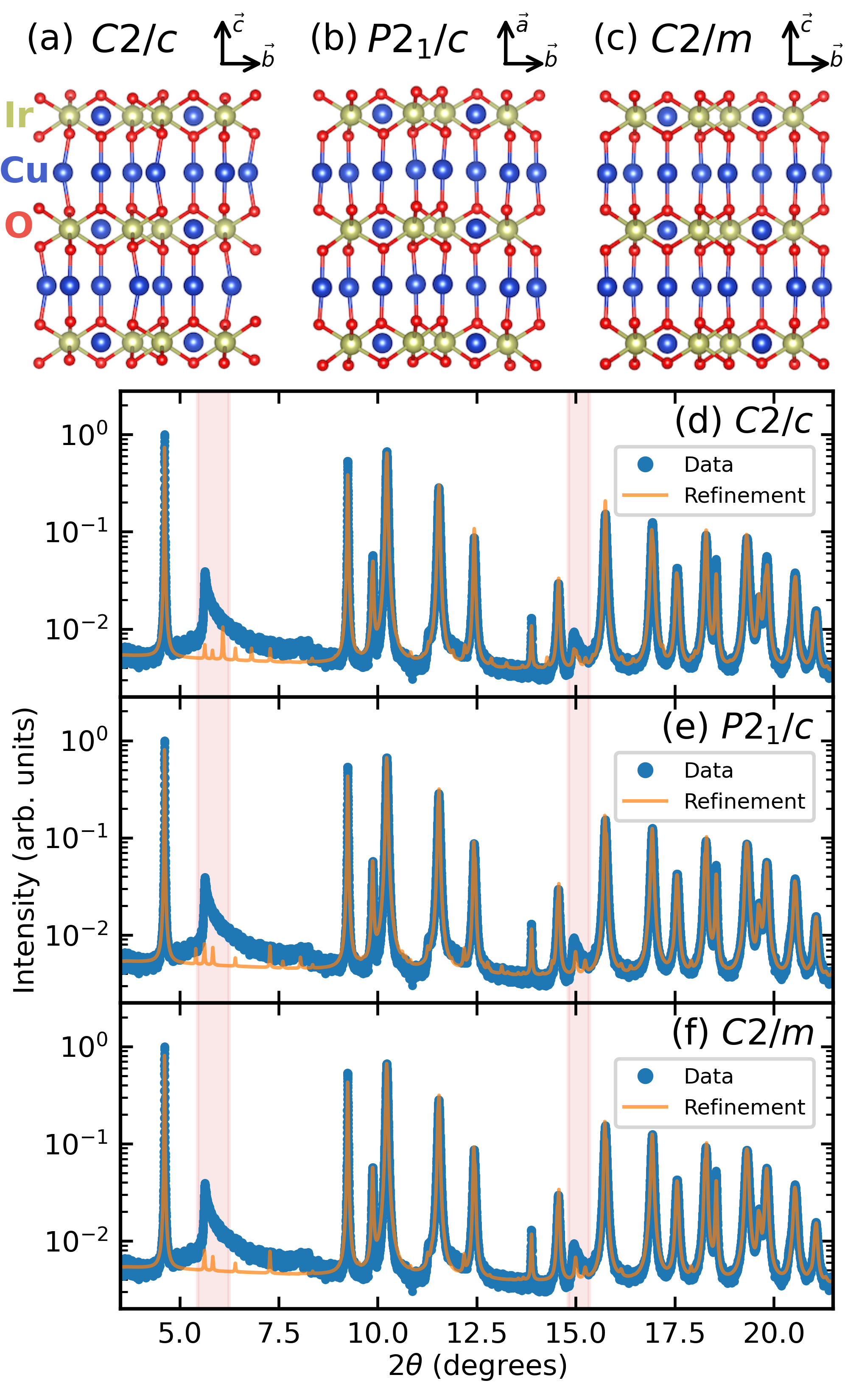}
\caption{\cuiro\ at ambient pressure (a) Previously suggested $C2/c$ structure \cite{Abramchuk2017}. (b) $P2_1/c$ structure obtained from evolutionary searches. (c) $C2/m$ structure obtained with DFT by relaxing the $C2/c$ structure. Ir, Cu, and O ions are represented in yellow, blue, and red, respectively. Note that in $P2_1/c$, the $a$ axis is out of the honeycomb plane. (d-f) Ambient pressure PXRD Rietveld refinement using the $C2/c$, $P2_1/c$, and $C2/m$ structures, respectively. The shaded areas mark the peaks that display Warren line shape asymmetric broadening and were excluded from the Rietveld refinement.}
\label{xrdphase1}
\end{figure}

\subsection{Phase 1: Ambient pressure structure revisited\label{phase1}}

\cuiro\ was originally solved with a $C2/c$ (mS48) structure comprising Ir-Cu layers with the signature Ir honeycomb framework linked by O-Cu-O dumbbells [Fig. \ref{xrdphase1}(a)] \cite{Abramchuk2017}. Even though the honeycomb planes are chemically well ordered, the presence of stacking faults was simulated in the Rietveld refinement of the PXRD data via partial occupancies of the in-plane Ir and Cu sites \cite{Abramchuk2017}. Stacking disorder has been shown to be irrelevant for the $\mathrm{Na_2IrO_3}$ structural stability \cite{Choi2012}, we thus simulated the \cuiro\ with a fully ordered 48-atom unit cell in the DFT calculations \cite{Abramchuk2017}.

\begin{figure}
\includegraphics[width=8.5 cm]{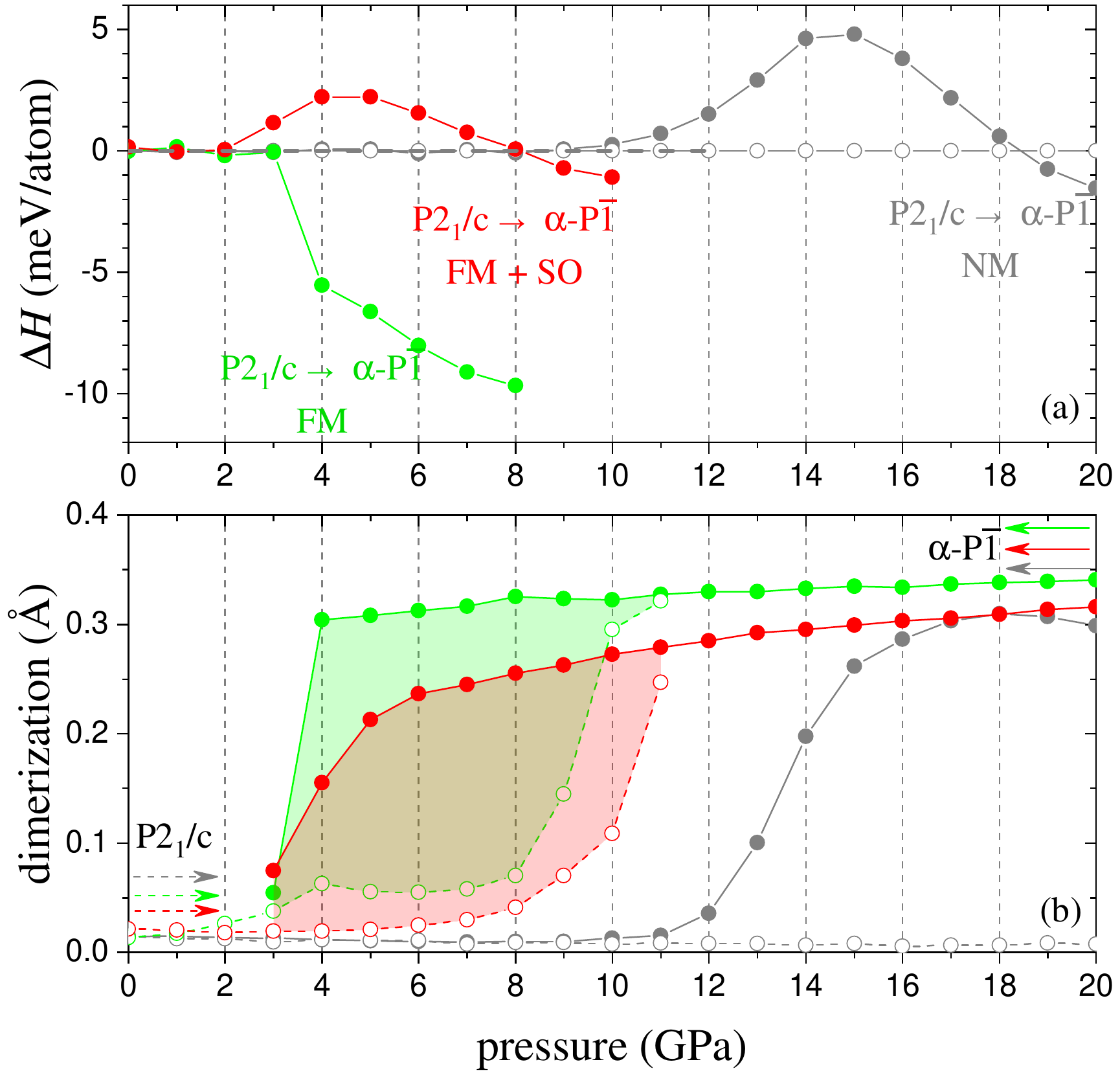}
\caption{Relative enthalpy (a) and dimerization (b) in two related $P2_1/c$ and $\alpha\text{-}P\overline{1}$ phases calculated with DFT+U with the NM, FM, and FM+SO settings. The local optimizations at each pressure were performed starting with either $P2_1/c$ relaxed at 0 GPa (hollow points) or $\alpha\text{-}P\overline{1}$ relaxed at 16 GPa (solid points). At the highest level of theory (red solid points), $\alpha\text{-}P\overline{1}$ becomes more stable than $P2_1/c$ at 8 GPa.}
\label{fig:enthDimer}
\end{figure}

As in the case of Na$_2$IrO$_3$ \cite{Choi2012}, $C2/c$ turned out not to be a local minimum for \cuiro\ and promptly relaxed into a simpler $C2/m$ (mS24) structure in our DFT+U+SO calculations. Figure \ref{xrdphase1}(a)\&(c) illustrate one of the clear differences between the two polymorphs: a smaller degree of the O-Cu-O bending in $C2/m$ (below 5 degrees) compared to that in $C2/c$ (up to 19 degrees). Unexpectedly, our evolutionary searches identified a new $P2_1/c$ (mP24) configuration with a similar morphology but significantly lower energy (7.2 meV/atom at 0 GPa) \cite{Supplemental}. In contrast to $C2/c$ and $C2/m$, it has a small in-and-out of plane modulation of the atomic positions in the pure Cu layer allowing a more uniform distribution of the Cu-Cu distances \cite{Supplemental}. The relative enthalpy also indicates that $P2_1/c$ remains favored under compression until 8 GPa (Fig. \ref{fig:enthDimer} and supplemental material \cite{Supplemental}). Having considered ferromagnetic (FM) and different antiferromagnetic (AFM) initial conditions in our DFT+U+SO calculations, we observed the FM ordering of Ir (0.15-0.35 $\mu_B$) and Cu ($\sim 0.75 \mu_B$) magnetic moments in the Ir-Cu layer to have the lowest enthalpy. This behavior differentiates Cu$_2$IrO$_3$ from Na$_2$IrO$_3$ and Li$_2$IrO$_3$ that display an AFM ordering at the DFT+U+SO level \cite{Singh2010,Choi2012}. Note, however, that a peak in the magnetic susceptibility of \cuiro\ is observed near 2~K, which is likely due to spin freezing of the $\mathrm{Cu^{2+}}$ ions, but that, combined with a negative Curie-Weiss temperature, may alternatively mark the onset of AFM order \cite{Abramchuk2017, Kenney2019}.

These structures feature similar PXRD patterns, with distinct differences only seen in certain weak peaks [Fig. \ref{xrdphase1}(d-f)]. However, the presence of stacking faults dramatically affects such peaks and makes it very difficult to uniquely determine the structure through PXRD. Rietveld refinement at ambient pressure yields a larger wR-factor for the $C2/c$ structure (16.5\%) than $C2/m$ or $P2_1/c$ (both 14.8\%), suggesting that the former is not the ground state structure, but also highlighting our inability to experimentally distinguish the $C2/m$ and $P2_1/c$ phases. In the remainder of this manuscript, we will use the theoretically determined $P2_1/c$ space group to describe phase 1.

\begin{figure}
\includegraphics[width = 8.25 cm]{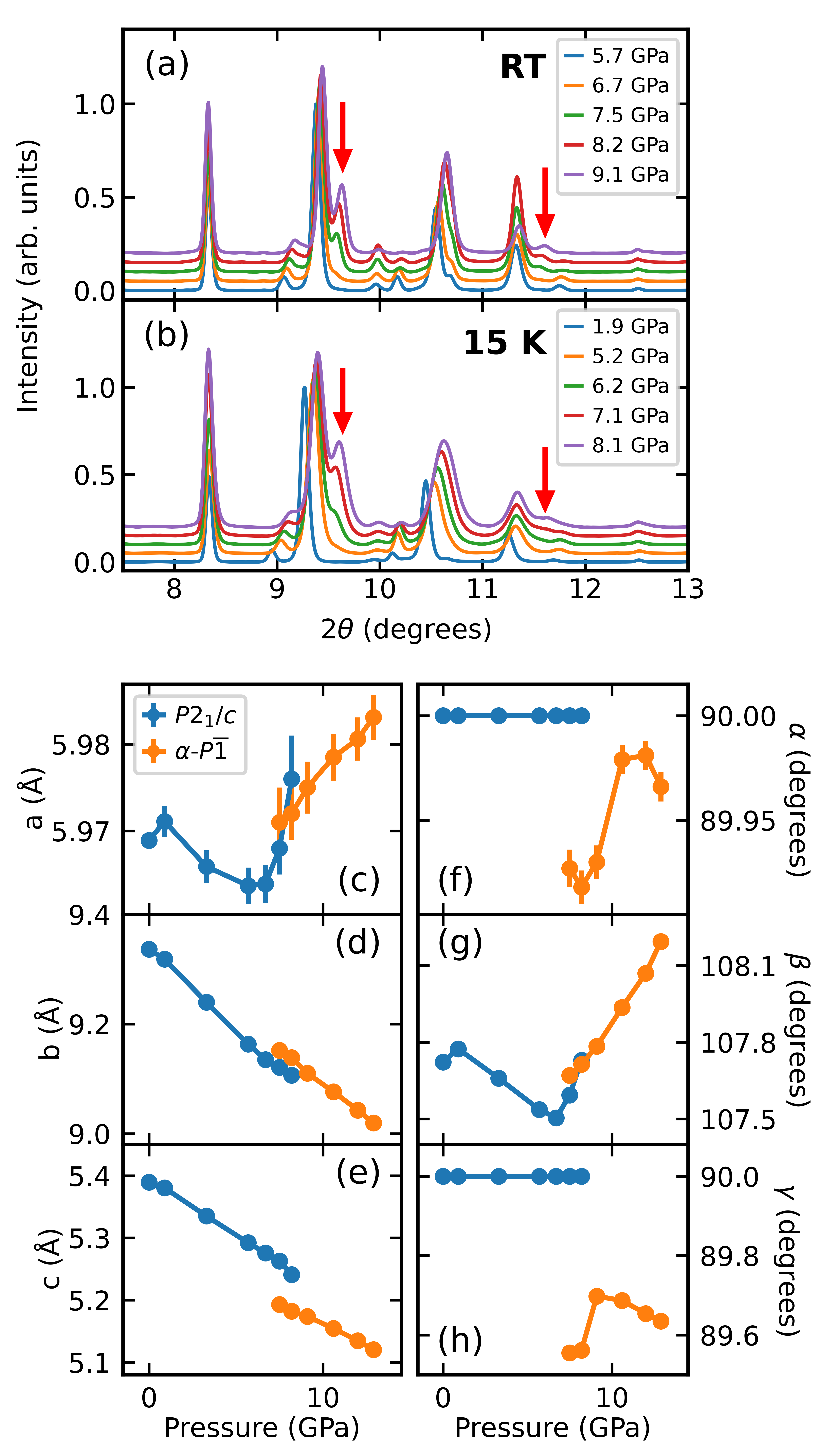}
\caption{(a)\&(b) \cuiro\ PXRD data across the phase 1 to 2 transition at RT and LT, respectively. The emergence of phase 2 is marked by a discontinuous reduction in one of the in-plane axis, which splits certain peaks, such as those marked in red arrows. (c-h) Pressure dependence of the lattice constants of \cuiro\ at room temperature. Note that in the $P2_1/c$ structure the $c$-axis lies in honeycomb plane. A mixture of phase 1 ($P2_1/c$) and 2 ($\alpha\text{-}P\overline{1}$) was used for 7.5 and 8.2 GPa. The changes in $\alpha$ and $\gamma$ for these pressures is likely due to correlation between the two phases.}
\label{xrdphase2}
\end{figure}

\subsection{Phase 2: Ir-Ir dimers \label{phase2}}

The onset of phase 2 occurs at $6 \pm 1.5$ GPa and $7.5 \pm 1$ GPa at LT and RT, respectively [Fig. \ref{xrdphase2}(a)\&(b)]. The $P2_1/c$ structure with a collapsed $c$ axis and slightly increased $b$ axis can reasonably reproduce the diffractogram [note that $b$ and $c$ are in-plane, Fig. \ref{xrdphase1}(b)], albeit with a sizable increase in wR (5.66\%, compared to $\lesssim 4\%$ at lower pressures \cite{Supplemental}), which suggests that phase 2 has a different space group.

Evolutionary searches at 16 GPa produced two alternative low-enthalpy $\alpha\text{-}P\overline{1}$ (aP24) and  $\beta\text{-}P\overline{1}$ (aP12) structures, with dimerized Ir-Ir frameworks, which are derived from the respective $P2_1/c$ and $C2/m$ phases with the same order in stability \cite{Supplemental}. The lower-enthalpy $\alpha\text{-}P\overline{1}$ is taken as the ground state crystal structure, being theoretically stable between 8-20 GPa.

\begin{figure}[t]
\includegraphics[width = 8.5 cm]{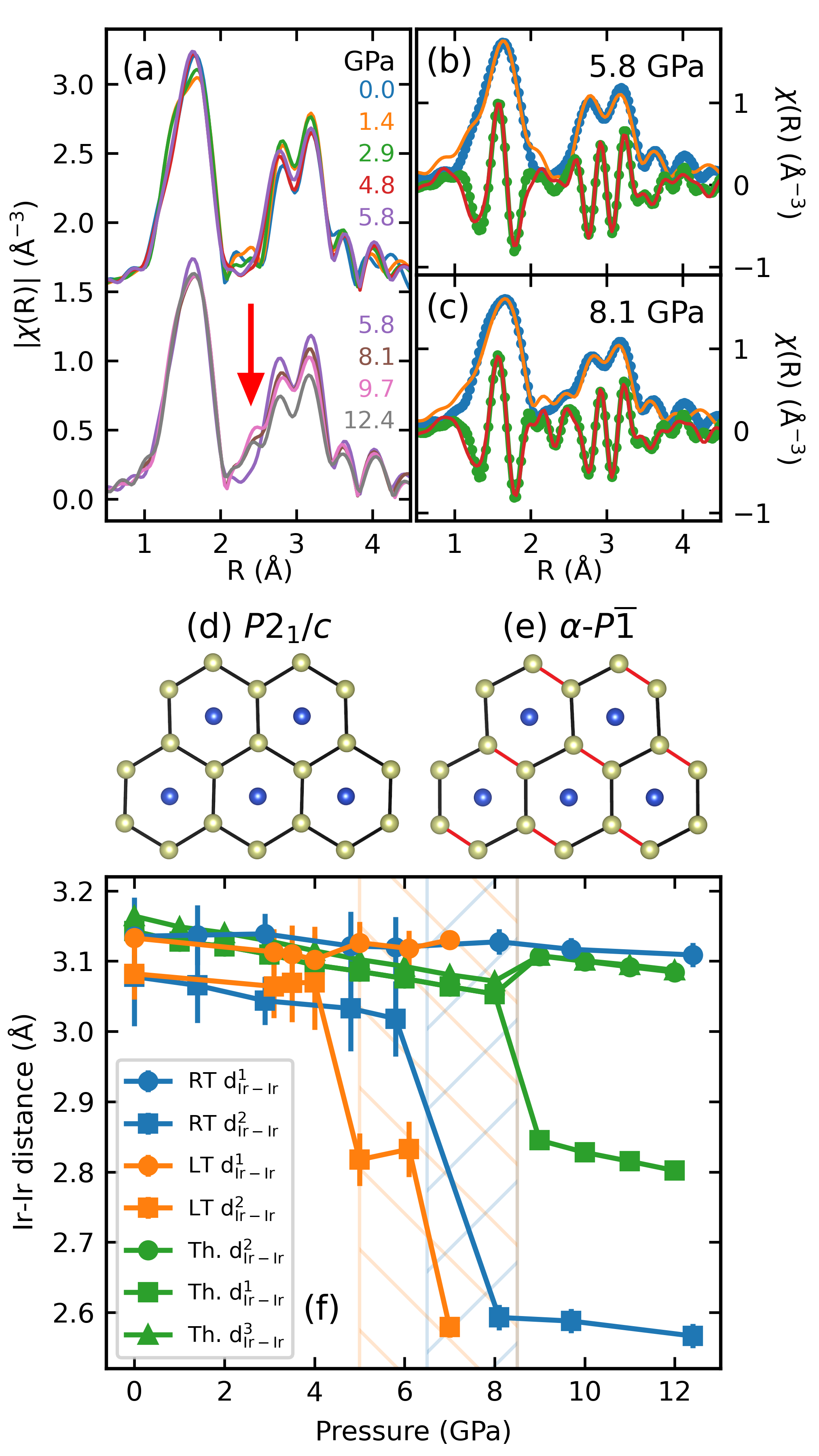}
\caption{\cuiro\ Ir $L_3$ XAFS. The $\chi(k)$ data were Fourier transformed using Hanning window with $k_{min} = 3.5~\mathrm{\AA}^{-1}$, $k_{max} = 13.5~\mathrm{\AA}^{-1}$, $dk = 1.0~\mathrm{\AA}^{-1}$, $k_{weight} = 2$. Fits were performed in $R$ space within 1.3-4.6 \AA. (a) Pressure dependence of the $\chi(R)$ magnitude. The dimerization transition is marked by an increase intensity around 2.5 $\mathrm{\AA}$ (red arrow). (b)\&(c) Results of the EXAFS modeling across the phase transition. (d)\&(e) Honeycomb structure of the $P2_1/c$ and $\alpha\text{-}P\overline{1}$ phases, with the Ir-Ir dimers shown in red. (f) Extracted pressure dependence of Ir-Ir distances at room temperature and 15 K. Note that the $P\overline{1}$ symmetry of phase 2 allows for three distinct Ir-Ir distances, but XAFS can resolve only two.}
\label{xafsphase2}
\end{figure}

Following the DFT prediction, we find that the $\alpha\text{-}P\overline{1}$ structure is a better match to the experimental data (wR = 4.96 \%), albeit with somewhat different lattice constants from those theoretically predicted \cite{Supplemental}. Extracted lattice constants are reported in Fig. \ref{xrdphase2}(c-h). Due to angular constraints and background from the diamond anvil cell environment, attempts to refine the atomic positions using the lower symmetry $\alpha\text{-}P\overline{1}$ space group have yielded inconsistent results, with large differences in Ir-Ir distances leading to small differences in wR.

\begin{figure}[b]
\includegraphics[width = 8.5 cm]{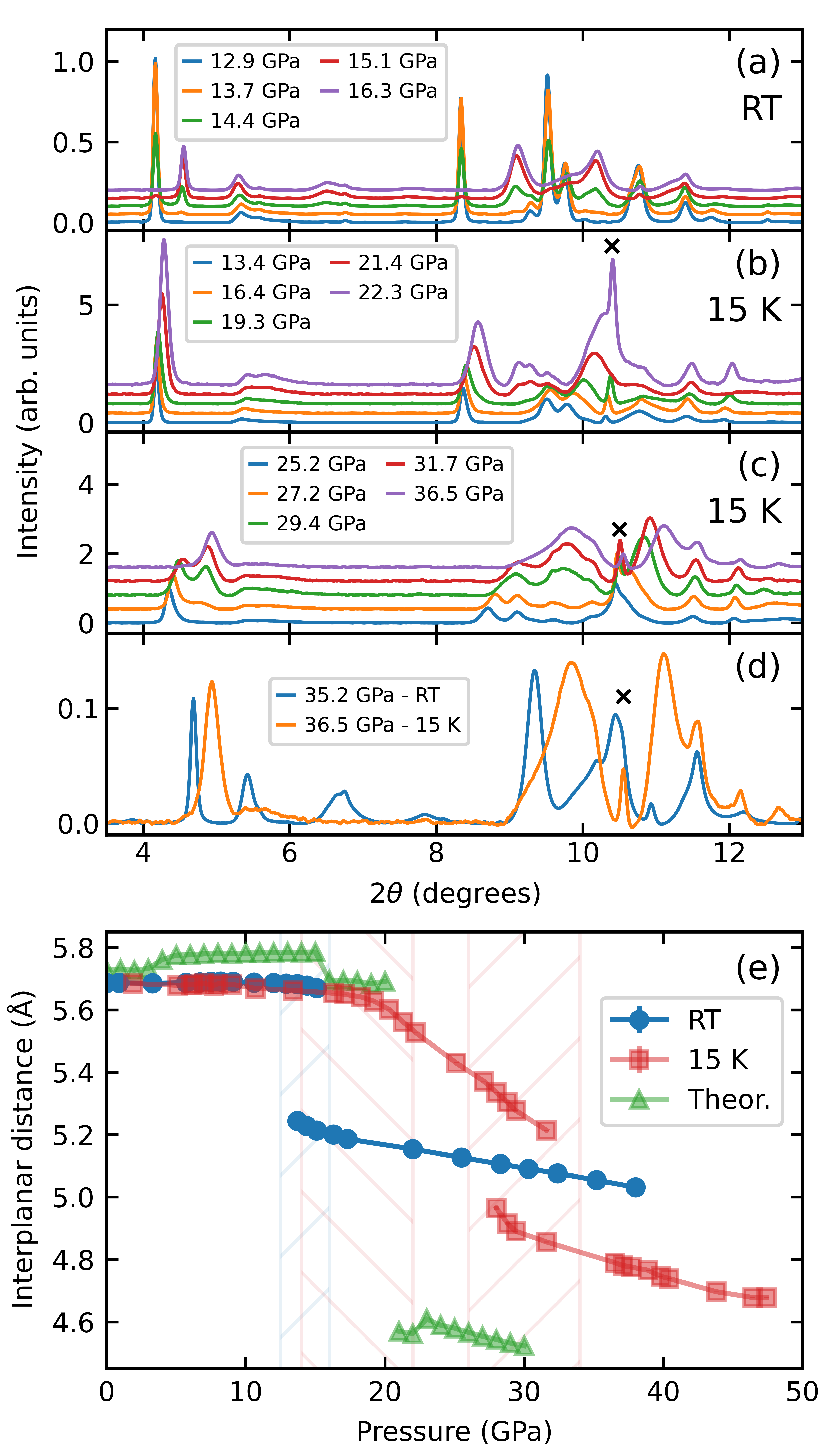}
\caption{\cuiro\ PXRD across phases 2-3 at RT (a), 2-4 at 15~K (b), and 4-5 at 15~K (c). (d) Comparison between the PXRD of phases 3 and 5, which occur at the same pressure, but are distinctly different. The cross symbol in (b-d) marks the position of the visible Au manometer Bragg peak. (e) Interplanar distance extracted from the (1 0 0) reflection [in $P2_1/c$ notation, peak around $2\theta \sim 4-5$ degrees in panels (a-d)] at room and low temperatures, as well as that in phases predicted using the evolutionary algorithm.}
\label{xrdphase3-5}
\end{figure}

XAFS measurements at the Ir $L_3$ edge were performed in order to experimentally probe for the presence of Ir-Ir dimers in phase 2. A distinct change in the XAFS data is observed across the phase 1 to 2 transition [Fig. \ref{xafsphase2}(a)]. The shift in spectral weight seen between 2-3.5~\AA\ points to a reconstruction of the shortest Ir-Cu and/or Ir-Ir distances. The XAFS of phase 1 is very well described by the $P2_1/c$ structure with nearly degenerate Ir-Ir distances. With the onset of phase 2, the XAFS signal can only be reproduced if one Ir-Ir distance is substantially shortened to $\mathrm{d_{Ir-Ir}^{short} \approx 2.6~\AA}$, demonstrating the presence of Ir-Ir dimers in phase 2. The dimerized bond length is similar to that observed in $\mathrm{Li_2IrO_3}$ \cite{Hermann2018, Veiga2019}, but shorter than predicted [$\sim 2.8~\mathrm{\AA}$, Fig. \ref{xafsphase2}(f)]. While the origin for such discrepancy is unclear, the predicted onset pressure is very close to the experimental value, and the resulting $\alpha\text{-}P\overline{1}$ phase is not only a better match to the PXRD data, but consistent with the symmetry seen in $\alpha\text{-}\mathrm{Li_2IrO_3}$ \cite{Hermann2018, Clancy2018}.

\subsection{Phases 3, 4 and 5: collapse of interplanar distance}

While Ir-Ir dimerization happens within the honeycomb plane in phase 2, the interplanar distance ($\mathrm{d_{planes}}$) is pressure independent until $\sim$ 15/20 GPa at RT/LT, respectively [Fig. \ref{xrdphase3-5} (e)]. This is a remarkable observation in itself, since one might expect the O-Cu-O dumbbells to be prone to buckling at lower pressures. At room temperature, the onset of phase 3 at $14.5 \pm 2$ GPa features a discontinuous reduction in the interplanar distance [Fig. \ref{xrdphase3-5}(a)\&(e)]. On the other hand, at low temperature, the onset of phase 4 at $18 \pm 4$ GPa is marked by a change in the $\mathrm{\partial d_{planes}/\partial P}$ [Fig. \ref{xrdphase3-5}(b)\&(e)], with a collapse of $\mathrm{d_{planes}}$ observed upon transition into phase 5 at $30 \pm 4$ GPa [Fig. \ref{xrdphase3-5}(c)\&(e)]. 

Substantial peak broadening at high pressure hinders the identification of the structural symmetry of phases 3, 4 and 5. Structural searches find a new $\alpha^{\prime}\text{-}P\overline{1}$ (aP12) phase above 20 GPa with collapsed interplanar distance, and no Ir-Ir dimers \cite{Supplemental}. However, this structure appears to be inconsistent with the PXRD data \cite{Supplemental}, preventing us from extracting the lattice constants beyond phase 2. As discussed in section \ref{discussion:hp_phases}, the strong PXRD temperature dependence beyond 15~GPa complicates the comparison between experiment and theory at these pressures.

Interestingly, phases 3 and 5, both of which happen after an interplanar distance collapse, are not identical, as evidenced by their distinct diffraction patterns at a similar pressure [Fig. \ref{xrdphase3-5}(d)]. Furthermore, their interplanar distance is not only markedly different (5.031 \AA\ at RT and 4.766 \AA\ at LT around 39~GPa), but displays distinct compressibility: -0.0076 \AA/GPa and -0.0115~\AA/GPa at RT and LT, respectively. In phases 1, 2, 4, and 5 the peaks around $2\mathrm{\theta} = 5-6^{\circ}$ are broadened due to stacking faults \cite{Abramchuk2017}. However, well defined peaks appear in this region for phase 3 at RT [Fig. \ref{xrdphase3-5}(d)], suggesting that the interplanar collapse at RT leads to a stronger structural correlation between the honeycomb planes. It is unclear, however, why such correlation does not emerge at low temperatures, where the interplanar distance is even shorter.
 
\begin{figure*}[!ht]
\includegraphics[width=\textwidth]{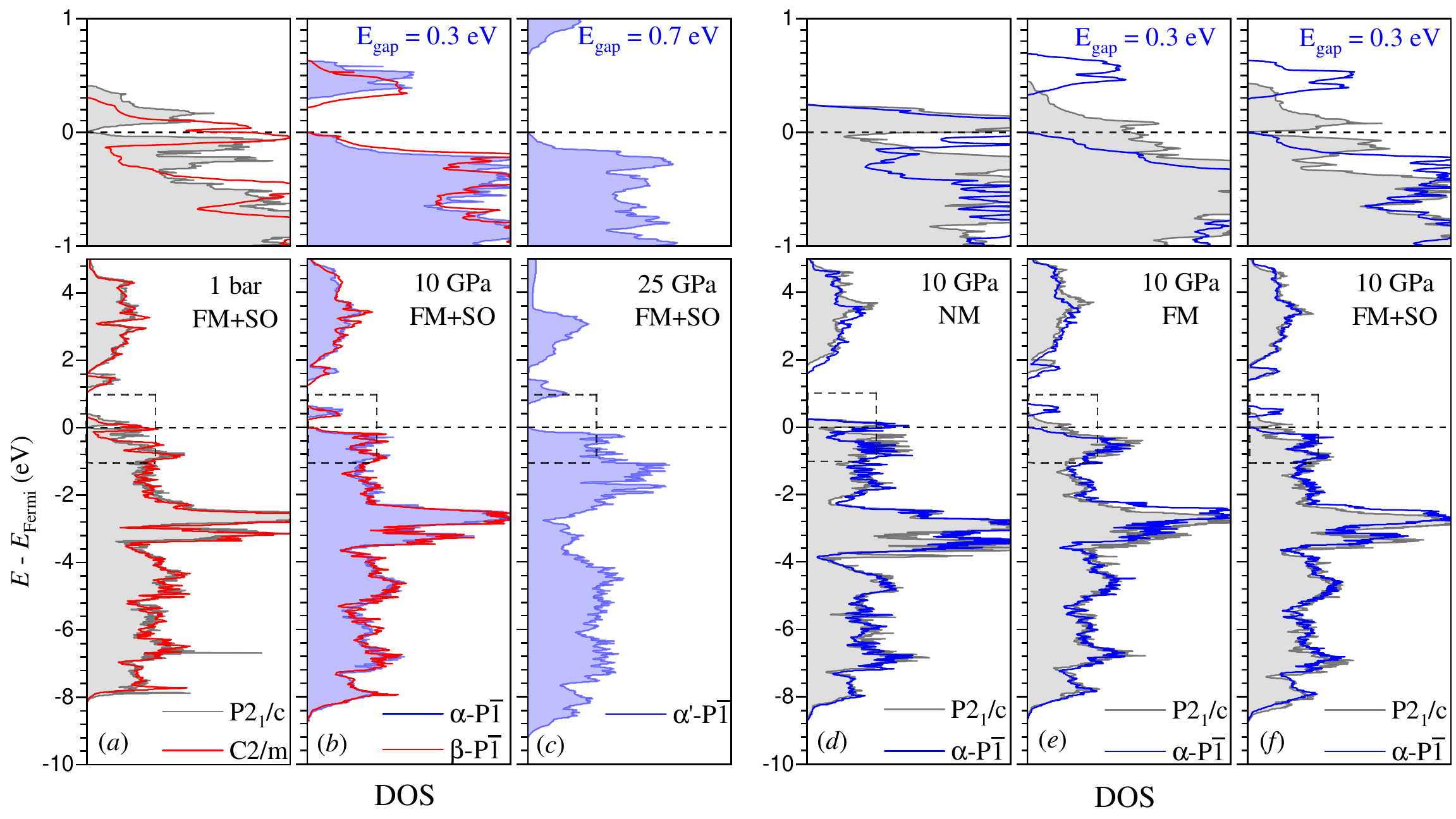}
\caption{Panels (a), (b), and (c) display the DOS at 1 bar ($P2_1/c$ and $C2/m$), 10 GPa ($\alpha\text{-}P\overline{1}$ and $\beta\text{-}P\overline{1}$), and 25 GPa ($\alpha^{\prime}\text{-}P\overline{1}$), respectively. Panels (d), (e), and (f) show the DOS of the $P2_1/c$ and $\alpha\text{-}P\overline{1}$ phases calculated with DFT+U with the NM, FM, and FM+SO settings. The region around the Fermi level is magnified above each panel. Decomposition of the DOS in (a-c) by element is shown in Figs. S5-7 \cite{Supplemental}.}
\label{fig:DOS}
\end{figure*}

\section{Discussion}

\subsection{Ir-Ir dimers in honeycomb iridates at high pressure}

Pressure-induced Ir-Ir dimers have been observed in $\mathrm{Li_2IrO_3}$ polymorphs around 2-4~GPa \cite{Veiga2017, Hermann2018, Clancy2018, Veiga2019}. However, dimerization is not seen experimentally in $\mathrm{Na_2IrO_3}$ up to at least 58~GPa \cite{Hermann2017, Layek2020}, although resistance measurements point to an electronic transition around 45~GPa, that is similar to the theoretically predicted dimerization in this material \cite{Hu2018}. Combined with the present work, the tendency for dimerization in $\mathrm{A_2IrO_3}$ compounds appears to correlate with the A-site ionic radius $R_{\mathrm{Li^{1+}}} \lesssim R_{\mathrm{Cu^{1+}}} < R_{\mathrm{Na^{1+}}}$ \cite{Shannon1976}. Future studies can verify this relationship by probing the recently synthesized $\mathrm{Ag_3LiIr_2O_6}$ and $\mathrm{H_3LiIr_2O_6}$ \cite{Kitagawa2018, Bahrami2019}.

In order to establish the factors promoting dimerization, we examined the $P2_1/c$ (phase 1) and $\alpha\text{-}P\overline{1}$ (phase 2) structures up to 20 GPa under different simulation conditions (Fig. \ref{fig:enthDimer}). We started the local optimizations with either non-dimerized $P2_1/c$ relaxed at 1 bar (hollow points) or $\alpha\text{-}P\overline{1}$ relaxed at 16 GPa (solid points) and used DFT+U calculations with non-magnetic (NM), ferromagnetic (FM), or FM plus spin orbit (FM+SO) interactions. The observed hysteresis in the pressure-induced dimerization transition indicates the existence of two local minima separated by a small barrier. Namely, the dimerized structure is more stable above 18, 4, and 8 GPa in the NM, FM, and FM+SO cases, respectively, but the transformations depend on the direction of the pressure change. For example, in the FM+SO local optimizations the structure fully dimerizes only around 11 GPa upon pressure increase and fully undimerizes around 2 GPa upon pressure decrease.

\subsection{Impact on the electronic structure}

Despite their similar morphology, the DOS of phases $P2_1/c$ and $C2/m$ at 1 bar are markedly different near the Fermi level [Fig. \ref{fig:DOS}(a)], with the former being a zero-gap semiconductor, which better matches the \cuiro\ insulating state \cite{Abramchuk2017}. Close inspection of the changes in electronic DOS with pressure for the different DFT settings provides insights into the stability ranges in Fig. \ref{fig:DOS} (d-f). Substantial differences in the transition pressure values are found in the NM, FM, and FM+SO calculations for $P2_1/c$ and $\alpha\text{-}P\overline{1}$ phases at 10 GPa (Fig. \ref{fig:enthDimer}). At the NM level, the DOS profile undergoes little change upon the Ir framework dimerization. With no apparent stability gain from the electronic state redistribution, the compound remains in the non-dimerized configuration up to at least 18 GPa. At the FM level, the dimerization has a pronounced effect on the DOS near the Fermi level turning the compound from a metal into a 0.3 eV semiconductor, similar to a Peierls transition. The transition is associated with the largest gain in enthalpy and, consequently, happens at the lowest pressure of 4 GPa. In the fully relativistic treatment, the SO coupling makes the material a zero-gap semiconductor even in the non-dimerized form, thus Cu$_2$IrO$_3$ does not stabilize nearly as much as in the FM-only case (without SO) when the dimerization opens up a similar 0.3 eV band gap. As a result, the transition occurs at a higher pressure (8 GPa) in excellent agreement with the measured value. Similar correlation between structural stability and (pseudo)gap formation have been reported in various materials \cite{Ravindran1997, Ravindran1998, AK09,AK31}.

The Ir-Ir dimerization in phase 2 ($\alpha\text{-}P\overline{1}$) drives the creation of $5d$ molecular orbitals, which are expected to destroy the $J_{\mathrm{eff}}$ state \cite{Veiga2017, Clancy2018, Takayama2019}. While the $J_{\mathrm{eff}}$ character of the $5d$ orbitals in phase 2 cannot be directly asserted here, the Ir projected DOS in this phase is characterized by an increased insulating gap \cite{Supplemental}, which is consistent with the presence of bonding and anti-bonding molecular orbitals. This result also agrees with the insulating behavior of $\mathrm{\beta\text{-}Li_2IrO_3}$ in its high-pressure dimerized phase \cite{Veiga2017}.

Beyond phase 2 ($>$ 15-20 GPa), the structure found in the evolutionary search ($\alpha^{\prime}\text{-}P\overline{1}$) appears to be inconsistent with the PXRD data \cite{Supplemental}. Theory successfully predicts the collapse of the interplanar distances above 15 GPa [Fig. \ref{xrdphase3-5}(e)], but the theoretical distance is substantially shorter than the experimental data. The electronic structure appears to be dramatically modified in this transition, with Ir-Cu hybridization largely dominating this phase \cite{Supplemental}. Notably, an increase in the DOS above the Fermi level is observed in one of the sites in the Cu plane, which suggests that Cu loses some electrons \cite{Supplemental}. This is not entirely surprising since the reduced interplanar distance likely increases the Cu coordination number, which favors the 2+ valence. Given the broad bands at this pressure, it is difficult to determine where these electrons go, but the oxygen orbitals are nominally filled, therefore, one can speculate that Ir may move towards 3+ valence ($5d^6$).

\subsection{Novel high-pressure phases \label{discussion:hp_phases}}

The collapsed interplanar distance beyond 15 GPa suggests that, structurally, \cuiro\ [$\mathrm{d_{planes} \sim 4.6-5.3 ~ \AA}$, Fig. \ref{xrdphase3-5}(e)] becomes more similar to the ambient pressure structure of \aliiro\ ($\mathrm{d_{planes} = 4.820 ~ \AA}$ \cite{OMalley2008}) and Na$_2$IrO$_3$ ($\mathrm{d_{planes} = 5.307 ~ \AA}$ \cite{Choi2012}). Notably, the theoretically predicted collapsed phase ($\alpha^{\prime}\text{-}P\overline{1}$) does not contain Ir-Ir dimers \cite{Supplemental}. However, the Ir and Cu orbitals are expected to largely hybridize at high pressure \cite{Supplemental}, leading to an electronic structure that is very different from the ambient pressure \aliiro\ or $\mathrm{Na_2IrO_3}$. Curiously, the $\alpha^{\prime}\text{-}P\overline{1}$ phase features a larger insulating gap compared to $P2_1/c$ or $\alpha\text{-}P\overline{1}$ [Fig. \ref{fig:DOS} (a-c)], despite the increased Ir-Cu hybridization and undimerized honeycomb layers \cite{Supplemental}. A similar collapse of the interplanar distance likely occurs in $\mathrm{H_2LiIr_2O_6}$ and $\mathrm{Ag_2LiIr_2O_6}$, which, as \cuiro, have intercalated honeycomb layers at ambient pressure \cite{Kitagawa2018, Bahrami2019}. While the Ag $4d$ orbital will likely hybridize with the Ir $5d$ in a similar manner as seen here, the high pressure electronic structure of a collapsed $\mathrm{H_2LiIr_2O_6}$ is less clear.

The presence of distinct high pressure phases at room and low temperature is not unusual, but the dramatic structural temperature dependence of \cuiro\ above 15 GPa is noteworthy. In particular, the collapse of the interplanar distance at 15 K requires twice (about 15 GPa) the pressure as at 300 K. Additionally, a significantly larger out-of-plane compressibility is seen at low temperature relative to RT. These observations may be explained by the presence of multiple nearly degenerate ground states. However, \cuiro\ could also be trapped in metastable phases at low temperature, which may involve temperature-dependent lattice/charge dynamics in the O-Cu-O dumbbells, possibly driven by 3d-5d hybridization and related oxidation of Cu sites as implied by the DFT calculations \cite{Supplemental}. Jahn-Teller modes associated with Cu$^{2+}$ ions may contribute to this unusual response. Both scenarios would largely hamper our ability to identify the observed phases via \emph{ab-initio} calculations, since the sizeable shear stress acting on the sample and the structure kinetics cannot be simply modeled. Directly probing the response of the electronic structure under pressure and variable temperature is required to elucidate if and how the unusual phase diagram of \cuiro\ is connected to underlying changes in electronic structure.

\section{Conclusions}

We studied the structure of \cuiro\ as a function of pressure and temperature using a combination of experimental and theoretical methods. Contrary to earlier reports, we suggest that the ambient pressure structure of \cuiro\ belongs to the new $P2_1/c$ space group. A transition into the $\alpha\text{-}P\overline{1}$ phase occurs around 7~GPa at both room temperature and 15~K. This phase is marked by the presence of Ir-Ir dimers that are expected to largely disrupt the Ir $J_{\mathrm{eff}} = 1/2$ state. To date, $\mathrm{Li_2IrO_3}$ was the only honeycomb iridate known to display dimers at high pressure \cite{Veiga2017, Hermann2018, Clancy2018, Veiga2019}, thus our results emphasize the likely universality of such structural motif in this system. Higher pressure not only drives \cuiro\ into novel structures, but also leads to dramatic differences in the structures at room and low temperatures. Ultimately, despite being different, the phases above 30~GPa are derived from a discontinuous reduction in the interplanar distance, which likely occurs due to the collapse of the O-Cu-O dumbbells. These results uncover a rich high pressure phase diagram for \cuiro, but they also raise further questions. In particular, the electronic structure across these multiple phase transitions remains to be experimentally explored. It is also unclear if the strong temperature dependence in the structural evolution under pressure is simply driven by kinetics (thermal energy), by unusual lattice/charge dynamics in the dumbbell structure, or if these are actually metastable phases that depend on the thermodynamic path. Finally, these results should stimulate high pressure work on $\mathrm{H_2LiIr_2O_6}$ and $\mathrm{Ag_2LiIr_2O_6}$ in order to determine the generality of the pressure-temperature phase diagram, and search for potential new phases.

\begin{acknowledgments}
We thank Curtis A. Kenney-Benson for the support during the x-ray diffraction measurements, and Larissa S. I. Veiga for critically reading the manuscript. Work at Argonne was supported by the U.S. DOE Office of Science, Office of Basic Energy Sciences, under Contract No. DE-AC02-06CH11357. Portions of this work were performed at HPCAT (Sector 16), Advanced Photon Source (APS), Argonne National Laboratory. HPCAT operations are supported by DOE-NNSA’s Office of Experimental Sciences. Helium and neon pressure media were loaded at GeoSoilEnviroCARS (The University of Chicago, Sector 13), Advanced Photon Source (APS), Argonne National Laboratory. GeoSoilEnviroCARS is supported by the National Science Foundation – Earth Sciences (EAR – 1634415) and Department of Energy- GeoSciences (DE-FG02-94ER14466). A.T. and A.N.K. acknowledge the NSF support (Award No. DMR-1821815) and the Extreme Science and Engineering Discovery Environment computational resources \cite{xsede} Award No. ACI-1548562, Project No. TG-PHY190024. We thank PRIUS for providing nanopolycrystalline diamond anvils.
\end{acknowledgments}

\bibliography{refs}

\end{document}